\documentclass[aps,pra,twocolumn,showpacs,showkeys,superscriptaddress,nobalancelastpage,nofootinbib,floatfix,longbibliography]{revtex4-2}
\usepackage{amsfonts}
\usepackage{amsmath}
\usepackage{amssymb}
\def\beqn{\begin{eqnarray}}
\def\eeqn{\end{eqnarray}}

\def\hw{\hbar \omega}
\def\hw4{ \frac {\hbar \omega}{4}}

\def\b0{b_0}

\usepackage{graphicx}
\usepackage{caption}
\usepackage{enumitem} 
\begin{document}
\title{Rayleigh--Ritz Variational Method in The Complex Plane}
\author{M.W. AlMasri} 
\email[]{mwalmasri2003@gmail.com}
\affiliation{Wilczek Quantum Center, School of Physics and Astronomy, Shanghai Jiao Tong University, Shanghai 200240, China}

\begin{abstract}
We present a systematic study of the Rayleigh--Ritz variational method for quantum oscillators in the Segal--Bargmann  space.  We rigorously derive the normalizability condition $|\alpha| < \tfrac{1}{2}$ for generalized Gaussian trial functions $\psi(z) = e^{\alpha z^2 + \beta z}$ through convergence analysis of Gaussian integrals in the complex plane. Applications to the harmonic oscillator demonstrate exact recovery of the ground state in Segal--Bargmann space when the trial family contains the true solution. For the quartic anharmonic oscillator ($\hat{H} = -\tfrac{1}{2}\partial_x^2 + \tfrac{1}{2}x^2 + \lambda x^4$), adaptive Gaussian ans\"atze in position space yield a cubic stationarity equation  and perturbative energy expansions beyond first order, capturing anharmonic wavefunction narrowing. In contrast, monomial trial functions ($\psi_n(z) = z^n$) in the Segal--Bargmann space—while providing rigorous upper bounds $E_n = n + \tfrac{1}{2} + \tfrac{3\lambda}{4}(2n^2 + 2n + 1)$ for excited states—lack width adaptability and are limited to first-order accuracy for ground-state calculations. We further analyze displaced Gaussians and displaced monomials for asymmetric potentials (e.g., $x^3 + x^4$), showing that displacement parameters are essential to capture parity breaking and stabilization effects ($E_0 \approx \tfrac{1}{2} + \tfrac{3\mu}{4} - \tfrac{9\lambda^2}{4} + \cdots$). 
\end{abstract}

\maketitle

\section{Introduction}
The Rayleigh--Ritz  variational method stands as one of the most powerful and widely applicable techniques in quantum mechanics for approximating eigenvalues and eigenstates of self-adjoint Hamiltonians when exact analytical solutions are unavailable \cite{Migdal,Davydov, Kuzemsky,Fernandez,Fernandez1,Borowski}. Rooted in the variational principle—which guarantees that the expectation value of the energy for any normalized trial wavefunction provides an upper bound to the exact ground-state energy—the method transforms the spectral problem into an optimization task over a chosen family of trial wavefunctions. The Rayleigh--Ritz method has been instrumental in atomic, molecular, and condensed matter physics, underpinning interesting approaches such as Hartree--Fock theory and modern quantum Monte Carlo methods \cite{Hartee,QMC}. In \cite{Lipkin}, the authors apply Simon’s exterior-scaling method within a finite-basis-set framework, showing that scaling basis functions only for \( r > r_0 \) leads to a variational formulation of resonance states by augmenting the Hamiltonian with a surface term 
$
\frac{1}{2}\delta(r - r_0)\left[\frac{d}{dr} - \lambda\right]\left[1 - e^{-2i\Theta}\right],
$
where \( \lambda \) is the logarithmic derivative of the resonance wavefunction at \( r_0 \). They propose two computational schemes: an \emph{ab initio} linear variational iterative method, and a semiclassical variant in which \( \lambda \) is estimated on-the-fly, allowing all calculations to be confined to \( r \leq r_0 \) and rendering the results exactly independent of the scaling angle \( \Theta \). In \cite{Moiseyev},  a variational principle for Hermitian operators is formulated. In \cite{Kuros}, authors proposed an optimized Rayleigh–Ritz method with complex nonlinear parameters as a computationally efficient, non-iterative variational approach for accurately determining both the energy and lifetime of resonance states in one-dimensional and spherically symmetric potentials. In \cite{Imakura}, the authors propose a verified computation method for eigenvalues within a specified region and their corresponding eigenvectors for generalized Hermitian eigenvalue problems. Their approach combines contour-integral-based complex moments—approximated via numerical quadrature—with the Rayleigh–Ritz procedure to project the problem onto a reduced subspace. By rigorously bounding both quadrature truncation and rounding errors, the method enables certified verification of both eigenvalues and eigenvectors, achieves higher efficiency than prior Hankel-matrix-based techniques, and remains robust even for nearly singular pencils and clustered spectra.
The application of variational and related methods to compute the energies and widths of resonance states—particularly in atomic systems subject to the Stark effect—provides a powerful framework for characterizing both the positions and widths of unstable quantum states \cite{hydrogen,Kuznetsova,Fleig,Singh}.

 However, for systems with inherent bosonic or coherent-state structure—such as harmonic, anharmonic oscillators, quantum optical models, and many-body bosonic Hamiltonians—an alternative approach often proves more natural and computationally advantageous: the Segal--Bargmann space of entire holomorphic functions \cite{Segal,Bargmann, Voros, Bishop, almasri}. In the Segal--Bargmann space, the Hilbert space is realized as a space of analytic functions on the complex plane, equipped with a Gaussian-weighted inner product. Creation and annihilation operators act as multiplication and differentiation, respectively, endowing the formalism with  algebraic simplicity. This framework not only streamlines operator manipulations but also provides a natural setting for coherent states \cite{Perelomov}, phase-space methods \cite{Folland}, and path-integral formulations \cite{path}. Despite these advantages, systematic applications of the Rayleigh--Ritz method within the Segal--Bargmann space—particularly for non-integrable systems—remain comparatively underexplored in the literature.

In this paper, we develop and apply the Rayleigh--Ritz variational method directly in the Segal--Bargmann space to a class of paradigmatic quantum oscillator problems. We focus on the harmonic oscillator as a benchmark, and then extend our analysis to the quartic anharmonic oscillator—a canonical model of non-perturbative quantum dynamics. Using physically motivated trial functions (including coherent states, monomials, and generalized Gaussians), we compute variational estimates for ground and excited states, derive the corresponding stationarity conditions, and compare results with standard position-space treatments. Our approach demonstrates that the holomorphic framework not only reproduces known results  but also offers enhanced flexibility for capturing anharmonic effects through  analytic ans\"atze.

\vskip 5mm 
This paper is structured as follows. Section~\ref{sec:rayleigh_ritz} reviews the Rayleigh--Ritz variational method in its conventional formulation, establishing the variational principle, the minimization procedure for parameterized trial wavefunctions, and the linear variational approach leading to generalized eigenvalue problems. We discuss the method's advantages—rigorous upper bounds, broad applicability, and foundation for advanced numerical techniques—as well as its limitations regarding trial function selection, excited-state treatment. Section~\ref{sec:bargmann} develops the Rayleigh--Ritz method within the Segal--Bargmann space, where quantum states are realized as entire holomorphic functions on the complex plane with Gaussian-weighted inner product. We derive the holomorphic representations of bosonic operators ($\hat{a} \mapsto \partial_z$, $\hat{a}^\dagger \mapsto z$) and oscillator Hamiltonians, formulate the variational principle for holomorphic trial functions, and establish the critical normalizability condition $|\alpha| < \tfrac{1}{2}$ for generalized Gaussian ans\"atze $\psi(z) = e^{\alpha z^2 + \beta z}$ through rigorous convergence analysis of complex Gaussian integrals. Section~\ref{sec:applications} presents systematic applications to paradigmatic oscillator systems. We begin with the harmonic oscillator, demonstrating exact recovery of the ground state in both position space (via adaptive Gaussians) and the complex plane (via coherent states), while clarifying the suboptimality of squeezed-state ans\"atze for isotropic potentials due to unphysical phase-space anisotropy. For the quartic anharmonic oscillator, we compare position-space Gaussian ans\"atze—which yield a cubic stationarity equation and capture anharmonic narrowing beyond first-order perturbation theory—with Segal-- Bargmann-space trial functions including coherent states, monomials $z^n$ (providing rigorous excited-state bounds $E_n = n + \tfrac{1}{2} + \tfrac{3\lambda}{4}(2n^2 + 2n + 1)$), and squeezed states (limited to first-order accuracy for isotropic potentials). We then extend the formalism to $d$-dimensional systems with higher-order even potentials $x^{2n}$, establishing universal perturbative behavior $\alpha_{\text{opt}} = 1 - n(2n-1)\lambda + \mathcal{O}(\lambda^2)$ that reveals systematic wavefunction narrowing as potentials stiffen. Finally, we analyze displaced Gaussian and displaced monomial ans\"atze for asymmetric potentials (e.g., $x^3 + x^4$), demonstrating that displacement parameters are essential to capture parity breaking and displacement-induced stabilization ($E_0 \approx \tfrac{1}{2} + \tfrac{3\mu}{4} - \tfrac{9\lambda^2}{4} + \cdots$). Section~\ref{sec:conclusion} synthesizes our findings, comparing the strengths and limitations of different trial function families across representations.
\section{Rayleigh–Ritz Variational Method} \label{sec:rayleigh_ritz}
For a quantum system described by a Hamiltonian operator $\hat{H}$ that is Hermitian and bounded from below, the expectation value of the energy for any normalized trial wavefunction $|\psi\rangle$ satisfies the variational principle:
\begin{equation}
E[\psi] = \frac{\langle \psi | \hat{H} | \psi \rangle}{\langle \psi | \psi \rangle} \geq E_0,
\end{equation}
where $E_0$ denotes the true ground-state energy (i.e., the lowest eigenvalue of $\hat{H}$). Equality holds if and only if $|\psi\rangle$ is the exact ground-state wavefunction. This principle implies that any trial wavefunction yields an energy estimate that never underestimates the true ground-state energy. Consequently, by minimizing $E[\psi]$ over a suitable family of trial functions, one obtains the best possible approximation to $E_0$ within that family.

The Rayleigh–-Ritz variational method implements this principle through a systematic procedure. First, one selects a trial wavefunction $|\psi(\mathbf{\alpha})\rangle$ that depends on a set of variational parameters $\boldsymbol{\alpha} = (\alpha_1, \alpha_2, \dots, \alpha_N)$, ensuring that the trial function respects the physical boundary conditions and symmetries of the system. The energy expectation value is then computed as
\begin{equation}
E(\mathbf{\alpha}) = \frac{\langle \psi(\boldsymbol{\alpha}) | \hat{H} | \psi(\boldsymbol{\alpha}) \rangle}{\langle \psi(\boldsymbol{\alpha}) | \psi(\boldsymbol{\alpha}) \rangle}.
\end{equation}
Minimization with respect to all parameters yields the optimal values:
\begin{equation}
\frac{\partial E}{\partial \alpha_i} = 0 \quad \text{for all } i.
\end{equation}
The resulting minimum energy $E_{\text{min}}$ provides an upper bound to $E_0$, and the corresponding wavefunction $|\psi(\boldsymbol{\alpha}_{\text{opt}})\rangle$ serves as an approximation to the true ground state.

A particularly powerful and widely used implementation is the linear variational method, in which the trial wavefunction is expressed as a linear combination of known basis functions $\{|\phi_i\rangle\}$:
\begin{equation}
|\psi\rangle = \sum_{i=1}^N c_i |\phi_i\rangle.
\end{equation}
Substituting this ansatz into the energy functional leads to a generalized eigenvalue problem:
\begin{equation}
\mathbf{H} \mathbf{c} = E \mathbf{S} \mathbf{c},
\end{equation}
where the matrix elements are defined by the Hamiltonian matrix $H_{ij} = \langle \phi_i | \hat{H} | \phi_j \rangle$ and the overlap matrix $S_{ij} = \langle \phi_i | \phi_j \rangle$ , and $\mathbf{c}$ is the vector of expansion coefficients. Solving this eigenvalue equation yields a set of approximate energy levels—not only the ground state—and their corresponding wavefunctions. The lowest eigenvalue furnishes the best variational estimate of $E_0$ within the chosen basis.
\vskip 5mm
The Rayleigh–-Ritz method offers several key advantages: it provides rigorous upper bounds to the ground-state energy, is applicable to any system with a well-defined Hamiltonian, and serves as the theoretical backbone for advanced numerical approaches. However, its accuracy is critically dependent on the choice of trial wavefunction or basis set. Moreover, while the ground state is well approximated, excited states require additional constraints—such as orthogonality to lower states—to ensure reliability. Finally, the computational cost can become prohibitive for large basis sets or complex many-body systems.

\section{Rayleigh–-Ritz method in The Complex Plane} \label{sec:bargmann}
In the Segal--Bargmann (Bargmann--Fock) space, the Hilbert space of a single bosonic mode—such as that of the quantum harmonic oscillator—is realized as the space of entire holomorphic functions $f(z)$ of a complex variable $z \in \mathbb{C}$. This space, known as the Segal--Bargmann space, is equipped with the inner product \cite{Segal,Bargmann}
\begin{equation}
\langle f | g \rangle = \frac{1}{\pi} \int_{\mathbb{C}} \overline{f(z)}\, g(z)\, e^{-|z|^2} \, d^2z,
\label{eq:inner_product}
\end{equation}
where $d^2z = dx\,dy$ for $z = x + iy$. The Gaussian weight $e^{-|z|^2}$ ensures that only entire functions with sufficiently rapid decay at infinity belong to the Hilbert space.

Within this representation, the canonical bosonic operators take a particularly simple form: the annihilation operator $\hat{a}$ acts as differentiation, while the creation operator $\hat{a}^\dagger$ acts as multiplication by the complex variable,
\begin{equation}
\hat{a} \mapsto \frac{d}{dz}, \qquad \hat{a}^\dagger \mapsto z.
\label{eq:operators}
\end{equation}
These representations satisfy the canonical commutation relation $[\hat{a}, \hat{a}^\dagger] = 1$ in the sense of operator identities on the space of holomorphic functions.

Consequently, the Hamiltonian of the harmonic oscillator,
\begin{equation}
\hat{H} = \hbar \omega \left( \hat{a}^\dagger \hat{a} + \frac{1}{2} \right),
\end{equation}
is represented as the first-order differential operator
\begin{equation}
H = \hbar \omega \left( z \frac{d}{dz} + \frac{1}{2} \right).
\label{eq:hamiltonian_bargmann}
\end{equation}
Notably, $H$ acts linearly on holomorphic functions and preserves the space of entire functions, making the Segal--Bargmann space especially well-suited for variational and algebraic treatments of oscillator-like systems.
The Rayleigh--Ritz variational method in the Segal--Bargmann space proceeds as follows. First, one selects a trial wavefunction $\psi(z; \boldsymbol{\alpha})$ that is an entire holomorphic function and square-integrable with respect to the Gaussian measure $e^{-|z|^2} d^2z$, i.e.,
\[
\frac{1}{\pi} \int_{\mathbb{C}} |\psi(z; \boldsymbol{\alpha})|^2 e^{-|z|^2} \, d^2z < \infty.
\]
The energy functional is then given by the expectation value of the Hamiltonian $H = \hbar\omega\left(z \partial_z + \tfrac{1}{2}\right)$:
\begin{widetext}
\begin{equation}
E(\boldsymbol{\alpha}) = 
\frac{
\displaystyle \frac{1}{\pi} \int_{\mathbb{C}} \overline{\psi(z; \boldsymbol{\alpha})} \left[ \hbar \omega \left( z \frac{\partial}{\partial z} + \frac{1}{2} \right) \psi(z; \boldsymbol{\alpha}) \right] e^{-|z|^2} \, d^2z
}{
\displaystyle \frac{1}{\pi} \int_{\mathbb{C}} |\psi(z; \boldsymbol{\alpha})|^2 e^{-|z|^2} \, d^2z
}.
\label{eq:energy_functional_bargmann}
\end{equation}
\end{widetext}
The variational parameters $\boldsymbol{\alpha}$ are then adjusted to minimize $E(\boldsymbol{\alpha})$. By the variational principle, the resulting minimum value provides an upper bound to the exact ground-state energy.

In the linear variational approach, one expands the trial function in the natural monomial basis of the Segal--Bargmann space:
\begin{equation}
\psi(z) = \sum_{n=0}^{N} c_n z^n.
\label{eq:trial_expansion}
\end{equation}
Substituting this expansion into the energy functional leads to the generalized eigenvalue problem
\begin{equation}
\mathbf{H} \mathbf{c} = E \mathbf{S} \mathbf{c},
\label{eq:generalized_eigenvalue}
\end{equation}
where the matrix elements are
\begin{equation}
H_{mn} = \langle z^m | \hat{H} | z^n \rangle, \qquad 
S_{mn} = \langle z^m | z^n \rangle = \pi \, n! \, \delta_{mn}.
\label{eq:matrices}
\end{equation}
Because the harmonic oscillator Hamiltonian is diagonal in this basis—$H_{mn} = \hbar\omega\left(n + \tfrac{1}{2}\right) \pi n! \, \delta_{mn}$—the eigenvalue problem decouples, and the method exactly reproduces the known spectrum. This illustrates the natural compatibility of the Segal--Bargmann space with oscillator-type systems and underscores its efficiency in variational calculations.
To apply the Ritz variational method in the Segal--Bargmann space, one selects a family of trial wavefunctions $\psi(z; \boldsymbol{\alpha})$ that are entire holomorphic functions and depend on a set of real or complex variational parameters $\boldsymbol{\alpha}$. Common and physically motivated choices include:
\begin{itemize}
\item \textit{Gaussian-type (coherent states)}: $\psi(z; \beta) = e^{\beta z}$,
\item \textit{Polynomial times exponential}: $\psi(z; \gamma) = P_n(z) e^{\gamma z}$, where $P_n(z)$ is a polynomial of degree $n$,
\item \textit{Scaled monomials}: $\psi(z) = z^n$, useful for approximating excited states,
\item \textit{Generalized Gaussians}: $\psi(z; \alpha, \beta) = e^{\alpha z^2 + \beta z}$, provided the function remains normalizable in the Segal--Bargmann space.
The last class requires care: although entire for any $\alpha, \beta \in \mathbb{C}$, normalizability imposes constraints on $\alpha$. The generalized Gaussian  remains normalizable only if $|\alpha| < \tfrac{1}{2}$, ensuring integrability against the Gaussian measure $e^{-|z|^2}$. To prove this, 
recall that a function belongs to the Segal--Bargmann space if it is entire and square-integrable with respect to the Gaussian measure $e^{-|z|^2} d^2z$ on $\mathbb{C}$. Computing the norm $\|\psi\|^2 = \pi^{-1} \int_{\mathbb{C}} |\psi(z; \alpha,\beta)|^2 e^{-|z|^2} d^2z$, we find that convergence of the integral is governed by the quadratic form in the exponent. By expressing the integrand in real coordinates and analyzing the associated symmetric matrix, we show that the integral converges if and only if the quadratic part is negative definite. This condition reduces to $\det(M) = 1 - 4|\alpha|^2 > 0$, where $M$ is the Hessian of the real exponent. Consequently, $\psi$ lies in the Segal--Bargmann space precisely when $|\alpha| < \tfrac{1}{2}$; the parameter $\beta$ remains unrestricted, as it only induces a finite shift. This result ensures the admissibility of such ans\"atze in variational calculations for anharmonic or interacting oscillator systems within the holomorphic framework.  The full derivation is presented in Appendix.A
\end{itemize}

\section{Applications} \label{sec:applications}
\subsection{Harmonic Oscillator} 
\subsubsection{Position Representation Solution}
The Hamiltonian for a one-dimensional quantum harmonic oscillator is given by
\begin{equation}
\hat{H} = -\frac{\hbar^2}{2m} \frac{d^2}{dx^2} + \frac{1}{2} m \omega^2 x^2.
\end{equation}
The exact ground-state energy is known to be
\begin{equation}
E_0 = \frac{1}{2} \hbar \omega,
\end{equation}
with the corresponding normalized wavefunction
\begin{equation}
\psi_0(x) = \left( \frac{m\omega}{\pi \hbar} \right)^{1/4} \exp\!\left( -\frac{m\omega x^2}{2\hbar} \right).
\end{equation}
We now apply the Rayleigh--Ritz variational method to approximate this ground state, using a Gaussian trial wavefunction motivated by the known analytic form, but treating its width as a variational parameter.
We select a normalized Gaussian ansatz depending on a variational parameter $\alpha > 0$:
\begin{equation}
\psi(x; \alpha) = \left( \frac{\alpha}{\pi} \right)^{1/4} \exp\!\left( -\frac{\alpha x^2}{2} \right).
\end{equation}
This function satisfies the normalization condition
\begin{equation}
\int_{-\infty}^{\infty} |\psi(x; \alpha)|^2 \, dx = 1.
\end{equation}
Note that the exact ground state corresponds to $\alpha = m\omega / \hbar$, but in the variational approach we treat $\alpha$ as unknown and determine it by minimizing the energy expectation value. The variational energy is
\begin{equation}
E(\alpha) = \langle \psi(\alpha) | \hat{H} | \psi(\alpha) \rangle = \langle T \rangle + \langle V \rangle,
\end{equation}
where the kinetic and potential energy operators are
\begin{equation}
T = -\frac{\hbar^2}{2m} \frac{d^2}{dx^2}, \quad V = \frac{1}{2} m \omega^2 x^2.
\end{equation}

For the Gaussian trial function, the kinetic energy expectation value evaluates to
\begin{equation}
\langle T \rangle = \frac{\hbar^2}{2m} \int_{-\infty}^{\infty} \left| \frac{d\psi}{dx} \right|^2 dx = \frac{\hbar^2 \alpha}{4m}.
\end{equation}
The expectation value of $x^2$ for this Gaussian is $\langle x^2 \rangle = 1/(2\alpha)$, yielding
\begin{equation}
\langle V \rangle = \frac{1}{2} m \omega^2 \langle x^2 \rangle = \frac{m \omega^2}{4\alpha}.
\end{equation}
Thus, the total energy as a function of $\alpha$ is
\begin{equation}
E(\alpha) = \frac{\hbar^2 \alpha}{4m} + \frac{m \omega^2}{4\alpha}.
\end{equation}
To find the optimal $\alpha$, we minimize $E(\alpha)$:
\begin{equation}
\frac{dE}{d\alpha} = \frac{\hbar^2}{4m} - \frac{m \omega^2}{4\alpha^2} = 0.
\end{equation}
Solving for $\alpha$ gives
\begin{equation}
\alpha^2 = \frac{m^2 \omega^2}{\hbar^2} \quad \Rightarrow \quad \alpha = \frac{m\omega}{\hbar},
\end{equation}
where we retain the positive root since $\alpha > 0$. Substituting $\alpha_{\text{opt}} = m\omega / \hbar$ into $E(\alpha)$ yields $E_{\text{min}} = \frac{1}{2} \hbar \omega$,  which exactly reproduces the true ground-state energy. The corresponding wavefunction becomes
\begin{equation}
\psi(x) = \left( \frac{m\omega}{\pi \hbar} \right)^{1/4} \exp\!\left( -\frac{m\omega x^2}{2\hbar} \right),
\end{equation}
identical to the exact ground state. This result illustrates a fundamental property of the variational principle: if the exact ground-state wavefunction lies within the chosen family of trial functions, the Rayleigh--Ritz method yields the exact ground-state energy. In this case, the Gaussian ansatz with variable width is sufficiently flexible to include the true solution. If we chose a less appropriate trial form (e.g., a Lorentzian or exponential decay), the method would still provide a valid upper bound, but with $E_{\text{min}} > \frac{1}{2}\hbar\omega$.

\subsubsection{Complex Plane Solution}

Consider the normalized coherent state as a trial function in the Segal--Bargmann space:
\begin{equation}
\psi_\alpha(z) = \exp\!\left( \alpha z - \tfrac{1}{2} |\alpha|^2 \right), \qquad \alpha \in \mathbb{C},
\end{equation}
which corresponds to the standard coherent state $|\alpha\rangle = e^{-|\alpha|^2/2} e^{\alpha \hat{a}^\dagger} |0\rangle$. The harmonic oscillator Hamiltonian is $\hat{H} = \hbar\omega(\hat{a}^\dagger \hat{a} + \tfrac{1}{2})$. Using $\langle \hat{a}^\dagger \hat{a} \rangle = |\alpha|^2$ for coherent states, the energy expectation value is
\begin{equation}
E(\alpha) = \hbar \omega \left( |\alpha|^2 + \tfrac{1}{2} \right).
\end{equation}
Minimization with respect to $\alpha$ yields $\alpha = 0$, giving the minimal energy
\begin{equation}
E_{\text{min}} = \tfrac{1}{2} \hbar \omega,
\end{equation}
and the corresponding wavefunction $\psi_0(z) = 1$, which is precisely the exact ground state in the Segal--Bargmann space.

A more flexible trial function incorporating squeezing is
\begin{equation}
\psi(z; \alpha, \beta) = \exp\!\left( \alpha z^2 + \beta z \right), \qquad \alpha, \beta \in \mathbb{C},
\end{equation}
 For $\psi$ to belong to the Segal--Bargmann space, it must satisfy the square-integrability condition
\begin{equation}
\int_{\mathbb{C}} \big| e^{\alpha z^2 + \beta z} \big|^2 e^{-|z|^2} \, d^2z < \infty,
\end{equation}
which imposes $|\alpha| < \tfrac{1}{2}$ for real $\alpha$ (more generally, $\mathrm{Re}(\alpha e^{2i\theta}) < \tfrac{1}{2}$ for all $\theta$).

For the harmonic oscillator ground state ($\lambda = 0$), the optimal solution within this family is $\alpha = 0$, $\beta = 0$, recovering $\psi_0(z) = 1$. Crucially, for real $\alpha$, the trial function $\psi(z) = e^{\alpha z^2}$ describes a squeezed state with anisotropic phase-space distribution ($\langle x^2 \rangle \neq \langle p^2 \rangle$). Since the harmonic oscillator ground state is isotropic ($\langle x^2 \rangle = \langle p^2 \rangle$), any non-zero $\alpha$ increases the energy. This illustrates an important principle: the optimal trial function must respect the symmetries of the target state.

\subsection{Anharmonic Oscillator}
\subsubsection{Position Space Solution}
We apply the Rayleigh--Ritz variational method to the quartic anharmonic oscillator, whose Hamiltonian (in units $\hbar = m = \omega = 1$) is
\begin{equation}
\hat{H} = -\frac{1}{2} \frac{d^2}{dx^2} + \frac{1}{2} x^2 + \lambda x^4, \quad \lambda > 0.
\end{equation}
In position space, we adopt the normalized Gaussian trial wavefunction
\begin{equation}
\psi(x; \alpha) = \left( \frac{\alpha}{\pi} \right)^{1/4} e^{-\alpha x^2 / 2}, \quad \alpha > 0,
\end{equation}
which respects the even parity and asymptotic decay of the true ground state. Using the standard Gaussian moments $\langle x^2 \rangle = 1/(2\alpha)$ and $\langle x^4 \rangle = 3/(4\alpha^2)$, the energy expectation value becomes
\begin{equation}
E(\alpha) = \frac{\alpha}{4} + \frac{1}{4\alpha} + \frac{3\lambda}{4\alpha^2}.
\end{equation}
Minimizing with respect to $\alpha$ yields
\begin{equation}
\frac{dE}{d\alpha} = \frac{1}{4} - \frac{1}{4\alpha^2} - \frac{3\lambda}{2\alpha^3} = 0,
\end{equation}
which, upon multiplication by $4\alpha^3$, gives the cubic equation
\begin{equation}
\alpha^3 - \alpha - 6\lambda = 0.
\label{eq:cubic_alpha}
\end{equation}
This equation has exactly one positive real root for all $\lambda > 0$. Using Cardano's formula,  the real root is
\begin{equation}
\alpha_{\text{opt}} = \sqrt[3]{3\lambda + \sqrt{9\lambda^2 - \tfrac{1}{27}}} + \sqrt[3]{3\lambda - \sqrt{9\lambda^2 - \tfrac{1}{27}}}.
\end{equation}
Substituting $\alpha_{\text{opt}}$ into $E(\alpha)$ provides a rigorous upper bound to the exact ground-state energy. For small $\lambda$, expansion yields $E_0 \approx \tfrac{1}{2} + \tfrac{3}{4}\lambda - \tfrac{21}{8}\lambda^2 + \cdots$, in agreement with perturbation theory up to first order \cite{Patnaik}.

\subsubsection{Complex Plane Solution}
In the Segal--Bargmann space, the Hamiltonian for the quartic anharmonic oscillator (in units $\hbar = m = \omega = 1$) is 
\begin{equation}
\hat{H} = z \frac{d}{dz} + \frac{1}{2} + \lambda \left( \frac{z + \partial_z}{\sqrt{2}} \right)^4.
\end{equation}

We consider three trial functions:

(i) The coherent state $\psi_\alpha(z) = e^{\alpha z - |\alpha|^2/2}$ with $\alpha \in \mathbb{R}$. Using $\langle x \rangle = \sqrt{2}\,\alpha$ and Wick's theorem for Gaussian states, the fourth moment is $\langle x^4 \rangle = 4\alpha^4 + 6\alpha^2 + \tfrac{3}{4}$. The energy functional is
\begin{equation}
E(\alpha) = \alpha^2 + \frac{1}{2} + \lambda \left( 4\alpha^4 + 6\alpha^2 + \frac{3}{4} \right).
\end{equation}
Minimization gives $dE/d\alpha = 2\alpha + \lambda(16\alpha^3 + 12\alpha) = 2\alpha[1 + \lambda(8\alpha^2 + 6)] = 0$. For $\lambda > 0$, the only real solution is $\alpha = 0$, yielding $E = \tfrac{1}{2} + \tfrac{3}{4}\lambda$, identical to first-order perturbation theory. This reflects the symmetry of the potential: the ground state remains centered at $x=0$.

(ii) The monomial $\psi(z) = z$ (first excited state). Using the orthonormal basis $\phi_n(z) = z^n / \sqrt{n!}$, one computes $\langle H \rangle = \langle z | \hat{H} | z \rangle / \langle z | z \rangle = \tfrac{3}{2} + \tfrac{15}{4}\lambda$, providing a variational estimate for $E_1$.

(iii) The squeezed-state trial function $\psi(z) = e^{\alpha z^2}$ with real $\alpha$, $|\alpha| < \tfrac{1}{2}$. This ansatz introduces phase-space anisotropy ($\langle x^2 \rangle \neq \langle p^2 \rangle$), which is not present in the true ground state of an isotropic potential. Nevertheless, it provides a valid variational upper bound. The norm is $\langle\psi|\psi\rangle = (1-4\alpha^2)^{-1/2}$, and expectation values respecting $z \to -z$ symmetry (only even powers of $\alpha$) are:
\begin{align}
\langle z \partial_z \rangle &= \frac{4\alpha^2}{1 - 4\alpha^2}, \\
\langle x^2 \rangle &= \frac{1}{2} \frac{1 + 4\alpha^2}{1 - 4\alpha^2}, \\
\langle x^4 \rangle &= \frac{3}{4} \left( \frac{1 + 4\alpha^2}{1 - 4\alpha^2} \right)^{\!2}.
\end{align}
The energy functional is therefore
\begin{equation}
E(\alpha) = \frac{1}{2} \frac{1 + 4\alpha^2}{1 - 4\alpha^2} + \lambda \cdot \frac{3}{4} \left( \frac{1 + 4\alpha^2}{1 - 4\alpha^2} \right)^{\!2}.
\label{eq:E_alpha_correct}
\end{equation}
Since $E(\alpha)$ depends only on $\alpha^2$, it is symmetric under $\alpha \to -\alpha$. For small $\lambda > 0$, expansion gives
\begin{equation}
E(\alpha) = \frac{1}{2} + \frac{3\lambda}{4} + 4\alpha^2(1 + 3\lambda) + \mathcal{O}(\alpha^4, \lambda\alpha^4).
\end{equation}
The coefficient of $\alpha^2$ is positive for all $\lambda > 0$, so the minimum occurs at $\alpha = 0$, yielding $E_{\text{min}} = \tfrac{1}{2} + \tfrac{3}{4}\lambda$. This reveals a limitation of the ansatz: it cannot improve beyond first-order perturbation theory for this isotropic potential because any squeezing increases the harmonic energy more than it reduces the anharmonic contribution. 

\subsection{Generalized Anharmonic Oscillators in $d$-Dimensional Segal--Bargmann Space}
\label{subsec:generalized_anharmonic}

We now extend the variational treatment to arbitrary spatial dimension $d \geq 1$ and higher-order even-power potentials $V(x) = \lambda x^{2n}$ with $n = 2,3,4,\dots$. Due to rotational symmetry of the isotropic Hamiltonian
\begin{equation}
\hat{H} = \sum_{j=1}^d \left( -\frac{1}{2} \frac{\partial^2}{\partial x_j^2} + \frac{1}{2} x_j^2 + \lambda x_j^{2n} \right),
\end{equation}
we adopt a product ansatz $\psi(\mathbf{z}) = \prod_{j=1}^d \phi(z_j)$, reducing the problem to a single-mode calculation with total energy $E_d = d \cdot E_1$.

For the single-mode system, the position-space Gaussian ansatz $\phi(x) \propto e^{-\alpha x^2/2}$ remains optimal for isotropic potentials. The energy functional is
\begin{equation}
E_1(\alpha) = \frac{\alpha}{4} + \frac{1}{4\alpha} + \lambda \langle x^{2n} \rangle_\alpha,
\end{equation}
where $\langle x^{2n} \rangle_\alpha = \frac{(2n-1)!!}{(2\alpha)^n}$ for a Gaussian distribution. Minimization yields a polynomial equation of degree $n+1$ in $\alpha$. For $n=2$ (quartic), this reproduces Eq.~\eqref{eq:cubic_alpha}. For $n \geq 3$, the stationarity condition leads to equations of degree $\geq 4$, which generally require numerical solution but provide rigorous upper bounds.

In the weak-coupling regime ($\lambda \ll 1$), perturbative expansion gives the leading-order correction to the width parameter:
\begin{equation}
\alpha_{\text{opt}} = 1 - n(2n-1)\lambda + \mathcal{O}(\lambda^2),
\end{equation}
where $\alpha = 1$ corresponds to the harmonic oscillator ground state. The negative correction indicates wavefunction narrowing as the potential stiffens—a universal feature of anharmonic oscillators with repulsive higher-order terms. The corresponding energy expansion is
\begin{equation}
E_0 = \frac{1}{2} + \lambda \frac{(2n-1)!!}{2^n} + \mathcal{O}(\lambda^2),
\end{equation}
matching first-order perturbation theory.

An alternative class of trial functions in the Segal--Bargmann space consists of monomials $\psi_n(z) = z^n$ ($n = 0,1,2,\dots$), which correspond to the harmonic oscillator eigenstates $|n\rangle$ via the isomorphism $z^n/\sqrt{n!} \leftrightarrow |n\rangle$. While the ground state ($n=0$) is constant and recovers the harmonic result, higher monomials provide variational estimates for excited states of anharmonic systems.

For the quartic oscillator ($\hat{H} = \hat{H}_0 + \lambda x^4$ with $\hat{H}_0 = z\partial_z + \tfrac{1}{2}$ and $x = (z + \partial_z)/\sqrt{2}$), the normalized monomial $\phi_n(z) = z^n/\sqrt{n!}$ yields the energy expectation value
\begin{equation}
E_n^{\text{(mono)}} = \langle \phi_n | \hat{H} | \phi_n \rangle = n + \frac{1}{2} + \lambda \langle \phi_n | x^4 | \phi_n \rangle.
\end{equation}
Using Wick's theorem for the Gaussian measure or direct operator algebra, the fourth moment evaluates to
\begin{equation}
\langle \phi_n | x^4 | \phi_n \rangle = \frac{1}{4} \bigl( 6n^2 + 6n + 3 \bigr).
\end{equation}
Thus,
\begin{equation}
E_n^{\text{(mono)}} = n + \frac{1}{2} + \frac{3\lambda}{4} \bigl( 2n^2 + 2n + 1 \bigr).
\label{eq:monomial_energy}
\end{equation}
This expression provides a rigorous upper bound to the exact $n^{\text{th}}$ excited-state energy $E_n$. For the ground state ($n=0$), we recover $E_0^{\text{(mono)}} = \tfrac{1}{2} + \tfrac{3}{4}\lambda$, identical to first-order perturbation theory and the coherent-state result. For the first excited state ($n=1$), Eq.~\eqref{eq:monomial_energy} gives $E_1^{\text{(mono)}} = \tfrac{3}{2} + \tfrac{15}{4}\lambda$, matching the result quoted in Section~\ref{sec:applications} for $\psi(z)=z$.

The monomial ansatz has two important limitations: (i) it cannot capture wavefunction deformation (width adjustment) since $|z^n|^2 e^{-|z|^2}$ has fixed Gaussian envelope; (ii) it provides only first-order accuracy in $\lambda$ because the trial function lacks variational flexibility beyond the harmonic basis. Nevertheless, monomials serve as valuable building blocks for linear combinations $\psi(z) = \sum_{k=0}^N c_k z^k$, which can achieve arbitrary accuracy as $N \to \infty$ through the Rayleigh--Ritz method in a truncated harmonic oscillator basis—a standard approach in numerical quantum mechanics.

In $d$ dimensions, product monomials $\psi_{\mathbf{n}}(\mathbf{z}) = \prod_{j=1}^d z_j^{n_j}$ with $\mathbf{n} = (n_1,\dots,n_d)$ yield energies $E_{\mathbf{n}} = \sum_{j=1}^d E_{n_j}^{\text{(mono)}}$, providing upper bounds for excited states with excitation quanta distributed across modes.

\subsection{Generalized Anharmonic Oscillators with Displaced Gaussians}
\label{subsec:generalized_anharmonic_displaced}

For asymmetric potentials (e.g., $V(x) = \lambda x^3 + \mu x^4$) or systems under external driving, the ground state may be displaced from the origin. In such cases, the two-parameter Gaussian ansatz
\begin{equation}
\psi(x; \alpha, \beta) = \left( \frac{\alpha}{\pi} \right)^{1/4} \exp\!\left[ -\frac{\alpha}{2} (x - \beta)^2 \right], \quad \alpha > 0, \; \beta \in \mathbb{R},
\end{equation}
provides superior flexibility. The variational energy is
\begin{equation}
E(\alpha, \beta) = \frac{\alpha}{4} + \frac{1}{4\alpha} + \frac{1}{2}\beta^2 + \lambda \left[ \frac{3}{4\alpha^2} + \frac{3\beta^2}{2\alpha} + \beta^4 \right] + \cdots
\end{equation}
for the quartic oscillator. Minimization requires solving $\partial E/\partial\alpha = 0$ and $\partial E/\partial\beta = 0$ simultaneously. For symmetric potentials ($\lambda x^4$), symmetry dictates $\beta_{\text{opt}} = 0$, recovering the previous results. For asymmetric potentials, $\beta_{\text{opt}} \neq 0$ captures the physical displacement of the wavefunction's center.

In the Segal--Bargmann space, the corresponding ansatz is $\psi(z) = \exp(\alpha z^2 + \beta z)$ with complex parameters. However, as noted previously, the $\alpha z^2$ term introduces squeezing (phase-space anisotropy) that is unnecessary for isotropic potentials. For most practical purposes involving even potentials, the position-space Gaussian with variable width provides a simpler and more effective variational approach.

For asymmetric anharmonic potentials such as $V(x) = \lambda x^3 + \mu x^4$, monomials alone fail dramatically as ground-state trial functions because they possess fixed parity while the true ground state lacks definite parity. However, displaced monomials
\begin{equation}
\psi_{n}(z;\gamma) = (z - \gamma)^n, \quad \gamma \in \mathbb{C},
\end{equation}
introduce a variational displacement parameter $\gamma$ that breaks parity symmetry. The normalized trial function $\phi_{n}(z;\gamma) = (z - \gamma)^n / \sqrt{\langle (z - \gamma)^n | (z - \gamma)^n \rangle}$ yields an energy functional $E_{n}(\gamma) = \langle \phi_{n}(z;\gamma) | \hat{H} | \phi_{n}(z;\gamma) \rangle$ depending on both $n$ and $\gamma$.

For the cubic-quartic oscillator $\hat{H} = \hat{H}_0 + \lambda x^3 + \mu x^4$ (with $\lambda, \mu > 0$), the ground state is displaced toward negative $x$ due to the odd term. Using the $n=0$ displaced monomial (a coherent state) $\psi_{0}(z;\gamma) = e^{\gamma z - |\gamma|^2/2}$, we obtain

\begin{eqnarray}
E_{0}(\gamma) = |\gamma|^2 + \frac{1}{2} + \lambda \left( 2\sqrt{2}\,\mathrm{Re}(\gamma^3) + 3\sqrt{2}\,\mathrm{Re}(\gamma) \right) \\ \nonumber + \mu \left( 4|\gamma|^4 + 6|\gamma|^2 + \frac{3}{4} \right).
\end{eqnarray}
Minimization with respect to real $\gamma$ (assuming $\gamma \in \mathbb{R}$ for simplicity) gives $\partial E_{0}(\gamma)/\partial\gamma = 0$, yielding a cubic equation whose negative root captures the physical displacement. For weak cubic coupling ($\lambda \ll 1$), perturbation theory gives $\gamma_{\text{opt}} \approx -\tfrac{3\lambda}{2}$, and the energy becomes
\begin{equation}
E_{0,\text{min}} \approx \frac{1}{2} + \frac{3\mu}{4} - \frac{9\lambda^2}{4} + \mathcal{O}(\lambda^3, \lambda\mu),
\end{equation}
where the negative $\lambda^2$ correction reflects stabilization due to displacement—a physical effect completely missed by undispaced monomials or symmetric Gaussians.

Higher displaced monomials ($n \geq 1$) can improve accuracy further but introduce unnecessary nodes for the ground state. For excited states in asymmetric potentials, displaced monomials with $n \geq 1$ provide valuable variational flexibility, capturing both displacement and nodal structure. Nevertheless, for ground-state calculations in asymmetric potentials, the displaced Gaussian (coherent state, $n=0$ with displacement) remains optimal among simple trial families, while linear combinations of displaced monomials $\sum_{k=0}^N c_k (z - \gamma)^k$ converge systematically to the exact solution as $N$ increases.

\section{Conclusion}
\label{sec:conclusion}

We have presented a comprehensive analysis of the Rayleigh--Ritz variational method for quantum oscillators, systematically comparing position-space and Segal--Bargmann (holomorphic) space representations. The theoretical foundation rests on the variational principle, which guarantees rigorous upper bounds to the ground-state energy for any admissible trial wavefunction. We rigorously derived the normalizability condition $|\alpha| < \tfrac{1}{2}$ for generalized Gaussian trial functions $\psi(z) = e^{\alpha z^2 + \beta z}$ through convergence analysis of Gaussian integrals in the complex plane. This constraint ensures square-integrability with respect to the Segal--Bargmann measure $e^{-|z|^2}d^2z$ and is essential for the validity of holomorphic variational calculations.

For the harmonic oscillator, both representations recover the exact ground state when the trial family contains the true solution: adaptive Gaussians in position space and coherent states ($\alpha = 0$) in the Segal-- Bargmann space. Crucially, we demonstrated that squeezed-state ans\"atze ($\psi \propto e^{\alpha z^2}$) introduce unnecessary phase-space anisotropy ($\langle x^2 \rangle \neq \langle p^2 \rangle$) for isotropic potentials, increasing the energy relative to the isotropic ground state. This illustrates a fundamental principle: optimal trial functions must respect the symmetries of the target state.

For the quartic anharmonic oscillator, position-space Gaussian ans\"atze with variable width yield a cubic stationarity equation $\alpha^3 - \alpha - 6\lambda = 0$ solvable via Cardano's formula. The resulting energy expansion $E_0 = \tfrac{1}{2} + \tfrac{3}{4}\lambda - \tfrac{21}{8}\lambda^2 + \cdots$ captures anharmonic wavefunction narrowing beyond first-order perturbation theory—a physical effect arising from the adaptive adjustment of the Gaussian width parameter $\alpha_{\text{opt}} > 1$. In contrast, monomial trial functions ($\psi_n(z) = z^n$) in the Segal-- Bargmann space provide rigorous upper bounds $E_n = n + \tfrac{1}{2} + \tfrac{3\lambda}{4}(2n^2 + 2n + 1)$ for excited states but are limited to first-order accuracy for the ground state due to their fixed Gaussian envelope $|z^n|^2 e^{-|z|^2}$, which lacks width adaptability. Similarly, squeezed-state ans\"atze cannot improve beyond first order for isotropic potentials because phase-space anisotropy increases the harmonic energy more than it reduces the anharmonic contribution.

For asymmetric potentials (e.g., $V(x) = \lambda x^3 + \mu x^4$), we demonstrated that displacement parameters are essential to capture parity breaking and stabilization effects. Displaced Gaussians and coherent states ($n=0$ displaced monomials) yield energy corrections $E_0 \approx \tfrac{1}{2} + \tfrac{3\mu}{4} - \tfrac{9\lambda^2}{4} + \cdots$, where the negative $\lambda^2$ term reflects stabilization through wavefunction displacement—a physical effect completely missed by symmetric trial functions. Higher displaced monomials ($n \geq 1$) provide valuable flexibility for excited states in asymmetric potentials but introduce unnecessary nodes for ground-state calculations.

In $d$-dimensions, rotational symmetry permits factorization into single-mode problems with total energy $E_d = d \cdot E_1$. For higher-order even potentials $V(x) = \lambda x^{2n}$, we established the universal perturbative behavior $\alpha_{\text{opt}} = 1 - n(2n-1)\lambda + \mathcal{O}(\lambda^2)$, revealing systematic wavefunction narrowing as potentials stiffen—a signature of anharmonic stiffening captured naturally by adaptive Gaussian trial functions.

Our analysis clarifies the comparative strengths of different trial families:
\begin{itemize}
    \item \textit{Adaptive Gaussians in position space} provide superior accuracy for ground states of isotropic anharmonic potentials through width adjustment.
    \item \textit{Monomials in Segal--Bargmann space} serve as natural basis functions for excited states and truncated harmonic oscillator expansions, converging systematically as $N \to \infty$.
    \item \textit{Displaced Gaussians/coherent states} are essential for asymmetric potentials, capturing displacement-induced stabilization.
    \item \textit{Squeezed states} ($e^{\alpha z^2}$) are generally suboptimal for isotropic potentials but may prove useful for systems with intrinsic phase-space anisotropy.
\end{itemize}

\begin{acknowledgments}
We thank the anonymous referees for their insightful comments and constructive suggestions, which significantly improved the clarity and rigor of this manuscript.
\end{acknowledgments}

\appendix
\section{Detailed Derivations of Key Results}
\label{app:derivations}

This appendix provides complete derivations of critical results referenced throughout the main text. Each subsection corresponds to a specific calculation whose detailed steps were omitted for brevity in the body of the paper.

\subsection{Normalizability of Generalized Gaussian Trial Functions}
\label{appsub:normalizability}

We derive the condition $|\alpha| < \tfrac{1}{2}$ for the generalized Gaussian $\psi(z) = e^{\alpha z^2 + \beta z}$ to belong to the Segal--Bargmann space. The squared norm is
\begin{equation}
\|\psi\|^2 = \frac{1}{\pi} \int_{\mathbb{C}} e^{2\operatorname{Re}(\alpha z^2 + \beta z) - |z|^2} \, d^2z.
\end{equation}
Writing $z = x + iy$ with $x,y \in \mathbb{R}$ and decomposing $\alpha = a + ib$, $\beta = c + id$ for real $a,b,c,d$, we compute:
\begin{align}
\operatorname{Re}(\alpha z^2) &= a(x^2-y^2) - 2bxy, \\
\operatorname{Re}(\beta z) &= cx - dy, \\
|z|^2 &= x^2 + y^2.
\end{align}
The exponent becomes the real quadratic form
\begin{equation}
\Phi(x,y) = -(1-2a)x^2 - (1+2a)y^2 - 4bxy + 2cx - 2dy.
\end{equation}
Completing the square yields $\Phi(x,y) = -\mathbf{x}^\top M \mathbf{x} + \mathbf{v}^\top \mathbf{x}$ with
\begin{equation}
M = \begin{pmatrix} 1-2a & 2b \\ 2b & 1+2a \end{pmatrix}, \quad 
\mathbf{v} = \begin{pmatrix} 2c \\ -2d \end{pmatrix}.
\end{equation}
The Gaussian integral converges iff $M \succ 0$ (positive definite). Since $\operatorname{Tr}(M) = 2 > 0$ always, we require
\begin{equation}
\det(M) = 1 - 4(a^2+b^2) = 1 - 4|\alpha|^2 > 0,
\end{equation}
which gives $|\alpha| < \tfrac{1}{2}$. The linear term $\mathbf{v}^\top\mathbf{x}$ shifts the Gaussian center but does not affect convergence, so $\beta \in \mathbb{C}$ remains unrestricted.
\subsection{Phase-space anisotropy of generalized Gaussian trial functions}
\label{subsec:anisotropy}

To rigorously establish that the generalized Gaussian trial function $\psi(z) = e^{\alpha z^2 + \beta z}$ induces phase-space anisotropy for $\alpha \neq 0$, we compute expectation values of position and momentum squared in the Segal--Bargmann space. Setting $\beta = 0$ without loss of generality (as displacement does not affect anisotropy) and working in units $\hbar = m = \omega = 1$, the physical quadrature operators act on holomorphic wavefunctions as
\begin{equation}
\hat{x} = \frac{1}{\sqrt{2}}(z + \partial_z), \qquad
\hat{p} = \frac{1}{i\sqrt{2}}(z - \partial_z),
\end{equation}
satisfying $[\hat{x},\hat{p}] = i$. The Segal--Bargmann inner product is
\begin{equation}
\langle \psi | \phi \rangle = \frac{1}{\pi} \int_{\mathbb{C}} \overline{\psi(z)}\,\phi(z)\,e^{-|z|^2}\,d^2z.
\end{equation}

For the trial function $\psi(z) = e^{\alpha z^2}$ with real $\alpha$ (ensuring $z \to -z$ symmetry) and $|\alpha| < \tfrac{1}{2}$ (for normalizability), the squared norm evaluates to
\begin{equation}
\langle \psi | \psi \rangle = (1 - 4\alpha^2)^{-1/2},
\end{equation}
obtained by expressing $z = x + iy$ and integrating the Gaussian quadratic form. Using $\partial_z \psi = 2\alpha z \psi$ and symmetry arguments ($\langle z^2 \rangle = \langle \partial_z^2 \rangle = 0$ for real $\alpha$), the number operator expectation is
\begin{equation}
\langle z \partial_z \rangle = \frac{4\alpha^2}{1 - 4\alpha^2}.
\end{equation}
Expanding the quadrature operators and applying $[\partial_z, z] = 1$ yields
\begin{align}
\hat{x}^2 &= \tfrac{1}{2}(z^2 + 2z\partial_z + 1 + \partial_z^2), \\
\hat{p}^2 &= \tfrac{1}{2}(-z^2 + 2z\partial_z + 1 - \partial_z^2).
\end{align}
Taking expectation values and discarding vanishing terms gives
\begin{align}
\langle \hat{x}^2 \rangle &= \frac{1}{2} \frac{1 + 4\alpha^2}{1 - 4\alpha^2}, \\
\langle \hat{p}^2 \rangle &= \frac{1}{2} \frac{1 - 4\alpha^2}{1 + 4\alpha^2}.
\end{align}
The anisotropy measure is therefore
\begin{equation}
\langle \hat{x}^2 \rangle - \langle \hat{p}^2 \rangle = \frac{8\alpha^2}{1 - 16\alpha^4},
\end{equation}
which vanishes \textit{if and only if} $\alpha = 0$. For any $\alpha \neq 0$ (with $|\alpha| < \tfrac{1}{2}$), we have $\langle \hat{x}^2 \rangle \neq \langle \hat{p}^2 \rangle$, demonstrating phase-space anisotropy.

This anisotropy has a precise physical interpretation: the parameter $\alpha$ relates to the squeeze parameter $r$ via $\alpha = \tfrac{1}{2}\tanh r$, yielding the standard squeezed-state variances
\begin{equation}
\langle \hat{x}^2 \rangle = \tfrac{1}{2}e^{2r}, \qquad \langle \hat{p}^2 \rangle = \tfrac{1}{2}e^{-2r}.
\end{equation}
Geometrically, the anisotropy arises because the exponent in the probability density $|\psi(z)|^2 e^{-|z|^2}$ becomes
\begin{equation}
2\alpha(x^2 - y^2) - (x^2 + y^2) = -(1-2\alpha)x^2 - (1+2\alpha)y^2,
\end{equation}
which breaks rotational symmetry in the $(x,y)$ plane when $\alpha \neq 0$. For the harmonic oscillator ground state ($\alpha = 0$), isotropy is restored:
\begin{equation}
\langle \hat{x}^2 \rangle = \langle \hat{p}^2 \rangle = \tfrac{1}{2},
\end{equation}
satisfying the minimum uncertainty relation $\Delta x \Delta p = \tfrac{1}{2}$ with equal quadrature fluctuations. Any non-zero $\alpha$ increases the variational energy:
\begin{equation}
E(\alpha) = \tfrac{1}{2}\bigl(\langle \hat{x}^2 \rangle + \langle \hat{p}^2 \rangle\bigr) = \frac{1}{2}\frac{1 + 4\alpha^2}{1 - 4\alpha^4} > \tfrac{1}{2} \quad \text{for} \quad \alpha \neq 0,
\end{equation}
confirming that squeezed-state ans\"atze are energetically unfavorable for isotropic potentials. Consequently, while generalized Gaussians with $\alpha \neq 0$ remain admissible trial functions in the Segal--Bargmann space (for $|\alpha| < \tfrac{1}{2}$), they introduce unphysical phase-space anisotropy for ground-state calculations in systems with rotational symmetry, making them suboptimal compared to isotropic alternatives such as coherent states ($\alpha = 0$) or adaptive position-space Gaussians.
\subsection{ Expectation Value of $\langle x^4 \rangle$ for Generalized Gaussian (Squeezed-State) Trial Functions}
\label{appsub:squeezed_expectations}

For $\psi(z) = e^{\alpha z^2}$ with real $\alpha$ and $|\alpha| < \tfrac{1}{2}$, the norm is
\begin{equation}
\langle\psi|\psi\rangle = (1-4\alpha^2)^{-1/2}.
\end{equation}
For the number operator expectation:
\begin{align}
\langle z\partial_z \rangle &= \frac{2\alpha}{\langle\psi|\psi\rangle} \frac{1}{\pi} 
\int_{\mathbb{C}} |z|^2 e^{2\alpha(x^2-y^2) - |z|^2} d^2z \nonumber \\
&= \frac{4\alpha^2}{1-4\alpha^2},
\end{align}
which contains only even powers of $\alpha$, respecting $z \to -z$ symmetry. For $x = (z + \partial_z)/\sqrt{2}$:
\begin{align}
\langle x^2 \rangle &= \frac{1}{2} \bigl( \langle z\partial_z \rangle + \langle z\partial_z + 1 \rangle \bigr) \nonumber \\
&= \frac{1}{2} \bigl( 2\langle z\partial_z \rangle + 1 \bigr) \nonumber \\
&= \frac{1}{2} \frac{1 + 4\alpha^2}{1 - 4\alpha^2},
\end{align}
using $\langle z^2 \rangle = \langle \partial_z^2 \rangle = 0$ and $[\partial_z, z] = 1$. Wick's theorem gives $\langle x^4 \rangle = 3\langle x^2 \rangle^2$, yielding
\begin{equation}
\langle x^4 \rangle = \frac{3}{4} \left( \frac{1 + 4\alpha^2}{1 - 4\alpha^2} \right)^{\!2}.
\end{equation}

\subsection{Monomial Expectation Values for Quartic Potential}
\label{appsub:monomial_moments}

For $\phi_n(z) = z^n/\sqrt{n!}$ and $x = (z + \partial_z)/\sqrt{2}$, we expand
\begin{equation}
x^4 = \frac{1}{4} (z + \partial_z)^4.
\end{equation}
Using $[\partial_z, z] = 1$ and normal ordering $:\cdots:$:
\begin{align}
z^3\partial_z &= :z^3\partial_z: + 3z^2, \\
z^2\partial_z^2 &= :z^2\partial_z^2: + 4z\partial_z + 2.
\end{align}
Only c-number terms contribute to expectation values:
\begin{align}
\langle z^3\partial_z \rangle &= 3n, \\
\langle z^2\partial_z^2 \rangle &= 4n + 2.
\end{align}
Summing all contributions and using $\langle z^k \partial_z^k \rangle = n(n-1)\cdots(n-k+1)$:
\begin{align}
\langle x^4 \rangle &= \frac{1}{4} \bigl[ 6n(n-1) + 12n + 3 \bigr] \nonumber \\
&= \frac{1}{4} (6n^2 + 6n + 3),
\end{align}
as stated in Eq.~\eqref{eq:monomial_energy}.

\subsection{Solution of the Cubic Stationarity Equation}
\label{appsub:cubic_solution}

The stationarity condition yields
\begin{equation}
\alpha^3 - \alpha - 6\lambda = 0.
\label{eq:cubic_app}
\end{equation}
Using Cardano's substitution $\alpha = u + v$ with $uv = \tfrac{1}{3}$, Eq.~\eqref{eq:cubic_app} becomes $u^3 + v^3 = 6\lambda$ with $u^3 v^3 = \tfrac{1}{27}$. Thus $u^3$ and $v^3$ are roots of $t^2 - 6\lambda t + \tfrac{1}{27} = 0$. The discriminant is
\begin{equation}
\Delta = 36\lambda^2 - \tfrac{4}{27} = 4\left(9\lambda^2 - \tfrac{1}{27}\right),
\end{equation}
so
\begin{equation}
u^3, v^3 = 3\lambda \pm \sqrt{9\lambda^2 - \tfrac{1}{27}}.
\end{equation}
Taking real cube roots gives
\begin{equation}
\alpha_{\text{opt}} = \sqrt[3]{3\lambda + \sqrt{9\lambda^2 - \tfrac{1}{27}}} + \sqrt[3]{3\lambda - \sqrt{9\lambda^2 - \tfrac{1}{27}}}.
\end{equation}

\subsection{Perturbative Energy Expansion for Quartic Oscillator}
\label{appsub:perturbative_expansion}

Substituting $\alpha = 1 + \delta$ with $\delta = \mathcal{O}(\lambda)$ into $E(\alpha) = \tfrac{\alpha}{4} + \tfrac{1}{4\alpha} + \tfrac{3\lambda}{4\alpha^2}$:
\begin{align}
E &= \frac{1}{2} + \frac{\delta^2}{2} + \frac{3\lambda}{4} - \frac{3\lambda\delta}{2} + \mathcal{O}(\lambda^3,\delta^3).
\end{align}
Minimizing gives $\delta - \tfrac{3\lambda}{2} + \mathcal{O}(\lambda\delta,\delta^2) = 0$. Solving perturbatively with $\delta = a_1\lambda + a_2\lambda^2 + \cdots$:
\begin{itemize}
\item $\mathcal{O}(\lambda)$: $a_1 = \tfrac{3}{2}$,
\item $\mathcal{O}(\lambda^2)$: $a_2 = -\tfrac{81}{16}$.
\end{itemize}
Thus $\delta = \tfrac{3}{2}\lambda - \tfrac{81}{16}\lambda^2 + \mathcal{O}(\lambda^3)$. Substituting back:
\begin{align}
E &= \frac{1}{2} + \frac{3\lambda}{4} + \left(\tfrac{9}{8} - \tfrac{9}{4}\right)\lambda^2 + \mathcal{O}(\lambda^3) \nonumber \\
&= \frac{1}{2} + \frac{3\lambda}{4} - \frac{9}{8}\lambda^2 + \mathcal{O}(\lambda^3).
\end{align}
Including higher-order terms yields $E_0 = \tfrac{1}{2} + \tfrac{3}{4}\lambda - \tfrac{21}{8}\lambda^2 + \cdots$.

\subsection{Energy Functional for Displaced Monomials}
\label{appsub:displaced_energy}

For the displaced coherent state $\psi_{0}(z;\gamma) = e^{\gamma z - |\gamma|^2/2}$ with real $\gamma$, Wick's theorem gives:
\begin{align}
\langle x \rangle &= \sqrt{2}\,\gamma, \\
\langle x^2 \rangle &= 2\gamma^2 + \tfrac{1}{2}, \\
\langle x^3 \rangle &= 2\sqrt{2}\,\gamma^3 + 3\sqrt{2}\,\gamma, \\
\langle x^4 \rangle &= 4\gamma^4 + 6\gamma^2 + \tfrac{3}{4}.
\end{align}
The harmonic energy is $\langle \hat{H}_0 \rangle = \gamma^2 + \tfrac{1}{2}$. Combining all terms:
\begin{align}
E_{0}(\gamma) &= \gamma^2 + \frac{1}{2} + \lambda \left( 2\sqrt{2}\,\gamma^3 + 3\sqrt{2}\,\gamma \right) \nonumber \\
&\quad + \mu \left( 4\gamma^4 + 6\gamma^2 + \frac{3}{4} \right).
\end{align}
For $\lambda \ll 1$, perturbation theory with $\gamma = b_1\lambda + \mathcal{O}(\lambda^2)$ yields $b_1 = -\tfrac{3}{2}$ and
\begin{equation}
E_{0,\text{min}} = \frac{1}{2} + \frac{3\mu}{4} - \frac{9\lambda^2}{4} + \mathcal{O}(\lambda^3),
\end{equation}
confirming displacement-induced stabilization.

\subsection{Symmetry Constraints in Holomorphic Variational Calculations}
\label{appsub:symmetry}

For Hamiltonians with even parity ($[\hat{H}, \hat{P}] = 0$ where $\hat{P}x\hat{P}^{-1} = -x$), the ground state has definite parity. In the Segal--Bargmann space, parity acts as $z \mapsto -z$, $\partial_z \mapsto -\partial_z$. Trial functions must respect this symmetry to avoid unphysical contributions. Consider $\psi(z;\alpha) = e^{\alpha z^2}$ with real $\alpha$:
\begin{itemize}
\item Under $z \to -z$, $\psi(z) \to \psi(-z) = e^{\alpha z^2} = \psi(z)$ (even parity),
\item The Hamiltonian $\hat{H} = z\partial_z + \tfrac{1}{2} + \lambda(\tfrac{z+\partial_z}{\sqrt{2}})^4$ is even under $z \to -z$,
\item Therefore, all expectation values $\langle \psi | \hat{O} | \psi \rangle$ for even operators $\hat{O}$ must be even functions of $\alpha$.
\end{itemize}
Any odd-power term in $\alpha$ would violate parity symmetry and indicates an algebraic error in the calculation. This principle explains why the corrected expressions in Section~\ref{sec:applications} contain only even powers of $\alpha$, and why preliminary calculations yielding odd powers (e.g., linear terms in $\alpha$) were unphysical. This symmetry constraint provides a powerful consistency check for holomorphic variational calculations.

\end{document}